\documentclass[11pt]{article}

\usepackage[preprint]{acl}
\definecolor{audioscorerorange}{HTML}{EE7B22}
\definecolor{lalmblue}{HTML}{4F81BD}
\usepackage{times}
\usepackage{latexsym}

\usepackage{amsmath}
\usepackage{amssymb} 
\usepackage[T1]{fontenc}
\usepackage[utf8]{inputenc}

\usepackage{microtype}

\usepackage{inconsolata}

\usepackage{graphicx}
\usepackage{placeins}
\usepackage{float}

\title{AudioScorer: Learning Question-Aware Token Relevance for Large Audio Language Models}

\author{
  \textbf{Marwan Farag} \\
  Institute of Signal Processing \\
  and System Theory (ISS) \\
  University of Stuttgart \\
  Stuttgart, Germany \\
  \texttt{marwan.farag@iss.uni-stuttgart.de}
  \And
  \textbf{Bin Yang} \\
  Institute of Signal Processing \\
  and System Theory (ISS) \\
  University of Stuttgart \\
  Stuttgart, Germany \\
  \texttt{bin.yang@iss.uni-stuttgart.de}
}

\begin{document}
\maketitle
% Avoid stretching inter-paragraph and heading glue to equalize columns.
\raggedbottom
% Keep float-only columns top-aligned rather than vertically centered.
\makeatletter
\setlength{\@fptop}{0pt}
\makeatother
\begin{abstract}
Large Audio Language Models (LALMs) reason over speech, environmental sounds,
music, and other acoustic information. However, their audio encoders produce
dense token sequences that make long-form audio costly to process. Existing audio
pruning methods reduce this cost using heuristics such as attention,
representation similarity, or temporal sampling, but do not explicitly learn
which tokens are most useful for a given question. We propose AudioScorer, a
lightweight model that scores audio tokens according to the user question. It is trained by distilling the ranking of audio tokens under the frozen
LALM's answer-gradient saliency, together with downstream answer supervision. Compared
with other pruning methods, AudioScorer achieves the best or tied-best result
at $10\%$ token retention on $7$ of $10$ semantic and acoustic tasks,
including speech recognition, emotion recognition, and audio-scene
classification.
\end{abstract}

\section{Introduction}

Multimodal Large Language Models (MLLMs) extend LLM reasoning to modalities such as vision and audio~\cite{alayrac2022flamingo,li2023blip,liu2023visual,openai20234v}. Particularly, Large Audio Language Models (LALMs)~\cite{chu2023qwen,tang2024salmonn,ghosh2026audio,xu2025qwen3} encode audio into dense embedding sequences projected into the LLM input space. Since the number of audio tokens grows with duration, long recordings substantially increase decoder cost. For example, Qwen2-Audio~\cite{chu2024qwen2} processes about $25$ audio tokens
per second, yielding roughly $750$ tokens for a $30$-second clip. A given
question, however, may depend on only a fraction of them, motivating
task-aware token pruning.
\suppressfloats[t]
\begin{figure}[!t]
  \centering
  \includegraphics[width=\linewidth,trim=5bp 15bp 5bp 15bp,clip]{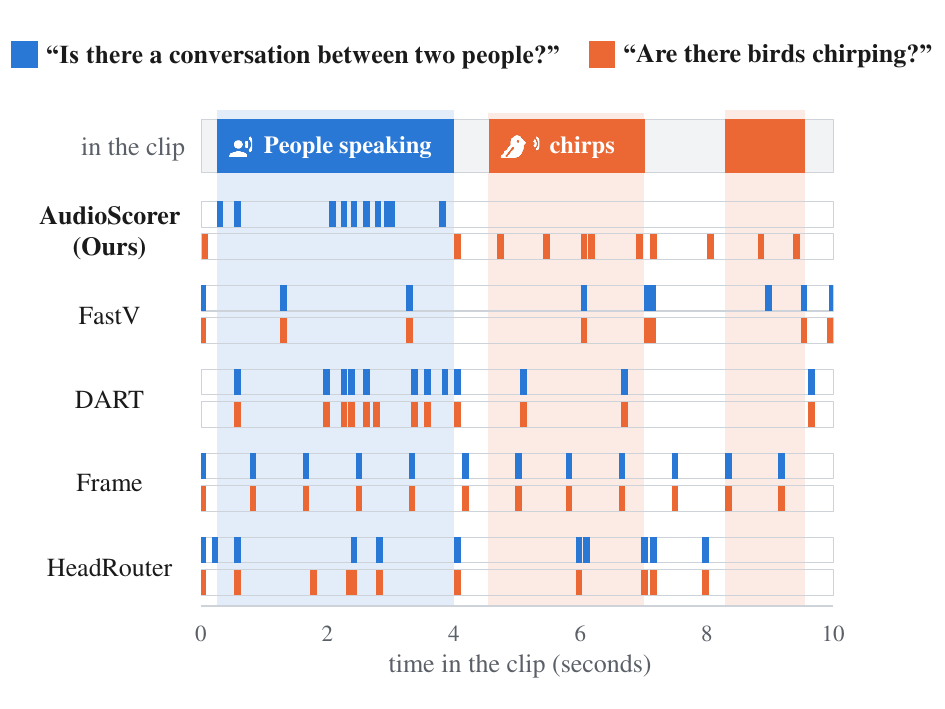}
      \caption{Question-dependent token selection for the same audio. AudioScorer adapts its selection to the question under a fixed budget. Audio and token-to-time mapping from \citet{ftar}.}

  \label{fig:question_switching}
\end{figure}

% Queue the full-width architecture while page one is being assembled.
% \begin{figure*}[!t]
%   \centering
%   \includegraphics[width=1.0\textwidth]{figures/scorer_architecture_acl.pdf}
% \caption{Overview of AudioScorer. Question-conditioned scores are produced using semantic and acoustic experts and trained with relevance distilled from the frozen LALM plus downstream answer supervision. At inference, these scores drive Top-$K$ pruning after the early decoder layers.}

%   \label{fig:scorer_architecture}
% \end{figure*}

\begin{figure}[t]
    \centering
    \includegraphics[width=0.80\columnwidth]{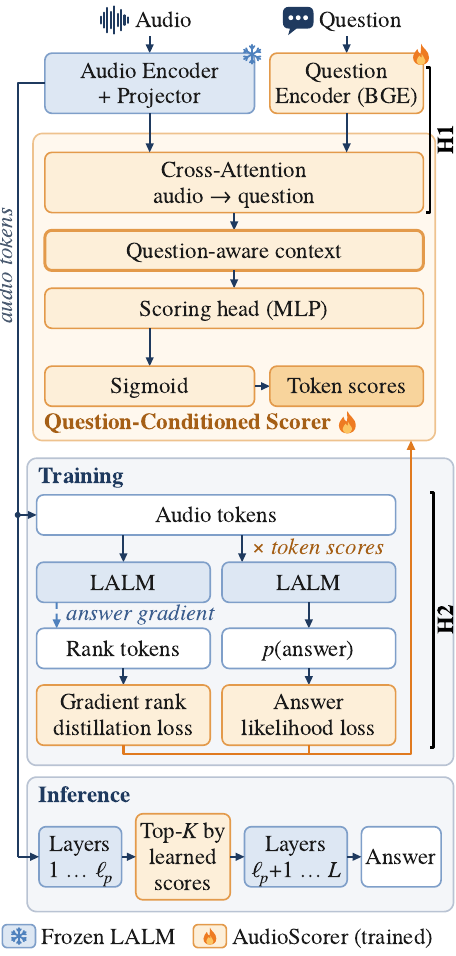}
     \caption{Overview of AudioScorer. It learns audio-token relevance from answer-gradient targets and downstream answer supervision. At inference, the scores drive Top-$K$ pruning after early decoder layers.}
    \label{fig:scorer_architecture}
\end{figure}

Existing pruning methods reduce token cost in MLLMs using signals such as attention~\cite{chen2024image,he2026headrouter,SparseVLM}, redundancy~\cite{wen2025stop}, temporal sampling~\cite{he2025audiomarathon}, or learned relevance. Question-aware audio pruning is also established~\cite{lin2025speechprune}, while learned relevance prediction has been explored in vision~\cite{takezoe2026learnpruner}. Existing question-aware pruning methods largely rely on predefined signals, while
learned relevance predictors are typically question-independent. Learning
audio-token utility jointly from the question and downstream answer quality
thus remains underexplored. This motivates our central question:
\emph{Can a scoring network learn question-conditioned audio-token
relevance from signals of downstream answer utility?}

We introduce AudioScorer, a lightweight question-conditioned scorer trained
by distilling the frozen LALM's gradient-based ranking of audio tokens
according to how strongly the correct answer depends on them, together with
downstream answer supervision. Figure~\ref{fig:question_switching} shows that AudioScorer's token selection
changes more with the question for the same audio clip than that of the other
pruning methods. At $10\%$ token retention on AudioMarathon~\cite{he2025audiomarathon}, AudioScorer is
best or tied-best on $7$ of $10$ tasks and leads on Speech Content, Audio
Classification, and Speaker Information. Notably, ASR reaches a $31.25\%$
word accuracy rate (WAR), compared with $5.15\%$ for the strongest baseline.
Our codebase will be released upon publication to support reproducibility.

\section{Method}

\label{sec:method}

\subsection{Design Hypotheses}
\label{sec:hypotheses}

Our goal is to retain the audio tokens most useful for answering the user's
question under a fixed token budget. We build AudioScorer around two
hypotheses.
\textbf{(H1) Token relevance depends on the question:} the same audio token
may be useful for one question and irrelevant for another.
\textbf{(H2) The frozen LALM can serve as a relevance teacher:} given the
question and reference answer, gradient-based sensitivity of the answer
likelihood provides a useful signal of audio-token relevance.

\subsection{AudioScorer Design}
\label{sec:scorer}

Figure~\ref{fig:scorer_architecture} illustrates
\textcolor{audioscorerorange}{AudioScorer} and its integration with the
\textcolor{lalmblue}{frozen LALM}. Its design follows directly from the two
hypotheses above.

\paragraph{Question-conditioned scoring (H1).}
We encode the question with \texttt{bge-base-en-v1.5}~\cite{bge_embedding}
and use a lightweight cross-attention layer in which each audio token attends
to the question tokens. A scoring head maps the resulting
question-conditioned representation to one score per audio token, allowing
the same recording to receive different rankings for different questions.

\begin{table*}[!t]
  \centering
  \footnotesize
  \setlength{\tabcolsep}{1.5pt}

  \resizebox{\textwidth}{!}{%
\begin{tabular}{ll|rrr|r||rrr|r||rrr|r||r||r}
    \hline
    & &
    \multicolumn{4}{c||}{\textbf{Speech Content}} &
    \multicolumn{4}{c||}{\textbf{Audio Classification}} &
    \multicolumn{4}{c||}{\textbf{Speaker Information}} &
    \textbf{Audio} &
    \textbf{Prefill} \\
    & &
    \multicolumn{4}{c||}{} &
    \multicolumn{4}{c||}{} &
    \multicolumn{4}{c||}{} &
    \textbf{auth.} &
    \textbf{compute} \\
    \cline{3-6}\cline{7-10}\cline{11-14}\cline{15-15}\cline{16-16}

    \textbf{Keep} & \textbf{Method}
    & ASR & SCR & SER & \textbf{Avg.}
    & SED & MC & ASC & \textbf{Avg.}
    & ER & SAR & SGR & \textbf{Avg.}
    & SD & TFLOPs \\
    \hline

    $100\%$ & Unpruned
    & 97.68 & 82.81 & 24.02 & 68.17
    & 70.18 & 100.00 & 70.31 & 80.16
    & 33.82 & 15.53 & 99.63 & 49.66
    & 90.11 & 46.04 \\
    \hline\hline

    & FastV
    & 97.03
    & 82.62
    & 24.52$^\dagger$
    & \underline{68.06}
    & 69.82
    & \textbf{100.00}
    & 70.62$^\dagger$
    & 80.15
    & 33.04
    & \underline{15.97}$^\dagger$
    & 98.98
    & 49.33
    & 87.53
    & 35.83 \\

    & SpeechPrune
    & 91.31
    & 80.51
    & 24.95$^\dagger$
    & 65.59
    & 68.59
    & \textbf{100.00}
    & 70.22
    & 79.60
    & \underline{35.14}$^\dagger$
    & 15.87$^\dagger$
    & 98.90
    & \underline{49.97}$^\dagger$
    & \textbf{93.93}$^\dagger$
    & 35.93 \\

    $70\%$ & DART
    & 95.93
    & 82.87$^\dagger$
    & \textbf{25.14}$^\dagger$
    & 67.98
    & \underline{69.84}
    & \textbf{100.00}
    & 70.50$^\dagger$
    & 80.11
    & \textbf{36.52}$^\dagger$
    & 15.90$^\dagger$
    & 99.19
    & \textbf{50.54}$^\dagger$
    & \underline{93.09}$^\dagger$
    & 35.88 \\

    & Frame
    & 95.28
    & 82.71
    & \underline{25.03}$^\dagger$
    & 67.67
    & \textbf{70.19}$^\dagger$
    & \textbf{100.00}
    & 70.93$^\dagger$
    & \textbf{80.37}$^\dagger$
    & 34.33$^\dagger$
    & 15.03
    & 98.98
    & 49.45
    & 85.03
    & 35.83 \\

    & HeadRouter
    & \underline{97.22}
    & \textbf{83.20}$^\dagger$
    & 22.74
    & 67.72
    & 69.31
    & \textbf{100.00}
    & \underline{70.99}$^\dagger$
    & 80.10
    & 31.82
    & \textbf{16.19}$^\dagger$
    & \underline{99.27}
    & 49.09
    & 80.92
    & 35.84 \\

    & AudioScorer
    & \textbf{97.49}
    & \underline{82.99}$^\dagger$
    & 25.00$^\dagger$
    & \textbf{68.49}$^\dagger$
    & 69.45
    & \textbf{100.00}
    & \textbf{71.03}$^\dagger$
    & \underline{80.16}
    & 34.67$^\dagger$
    & 15.09
    & \textbf{99.49}
    & 49.75$^\dagger$
    & 82.47
    & 35.99 \\
    \hline\hline

    & FastV
    & 92.26
    & \underline{82.53}
    & \textbf{25.31}$^\dagger$
    & \underline{66.70}
    & \underline{70.15}
    & \textbf{100.00}
    & \underline{71.20}$^\dagger$
    & \textbf{80.45}$^\dagger$
    & 35.27$^\dagger$
    & \underline{16.55}$^\dagger$
    & 98.18
    & 50.00$^\dagger$
    & 70.80
    & 26.27 \\

    & SpeechPrune
    & 68.26
    & 72.15
    & 22.47
    & 54.29
    & 67.92
    & \textbf{100.00}
    & 68.83
    & 78.92
    & \textbf{40.76}$^\dagger$
    & \textbf{17.88}$^\dagger$
    & 98.76
    & \textbf{52.47}$^\dagger$
    & \textbf{85.02}
    & 26.34 \\

    $40\%$ & DART
    & 85.21
    & 77.08
    & \underline{25.15}$^\dagger$
    & 62.48
    & 65.87
    & \textbf{100.00}
    & 70.44$^\dagger$
    & 78.77
    & \underline{37.98}$^\dagger$
    & 14.09
    & \underline{99.05}
    & \underline{50.37}$^\dagger$
    & \underline{79.22}
    & 26.32 \\

    & Frame
    & 86.47
    & 79.84
    & 25.11$^\dagger$
    & 63.81
    & 69.43
    & \textbf{100.00}
    & 71.17$^\dagger$
    & 80.20$^\dagger$
    & 32.78
    & 13.42
    & 98.62
    & 48.27
    & 72.79
    & 26.27 \\

    & HeadRouter
    & \underline{93.28}
    & 80.58
    & 24.52$^\dagger$
    & 66.13
    & \textbf{70.19}$^\dagger$
    & \textbf{100.00}
    & 71.07$^\dagger$
    & \underline{80.42}$^\dagger$
    & 27.11
    & 16.11$^\dagger$
    & 97.69
    & 46.97
    & 63.40
    & 26.28 \\

    & AudioScorer
    & \textbf{96.94}
    & \textbf{83.25}$^\dagger$
    & 23.85
    & \textbf{68.01}
    & 69.10
    & \textbf{100.00}
    & \textbf{71.78}$^\dagger$
    & 80.29$^\dagger$
    & 34.19$^\dagger$
    & 14.42
    & \textbf{99.19}
    & 49.27
    & 71.01
    & 26.43 \\
    \hline\hline

    & FastV
    & 0.00
    & \underline{69.20}
    & \underline{21.61}
    & \underline{30.27}
    & \textbf{70.62}$^\dagger$
    & \underline{99.26}
    & \underline{67.48}
    & \underline{79.12}
    & 31.25
    & \textbf{15.55}$^\dagger$
    & 94.29
    & 47.03
    & 35.72
    & 17.35 \\

    & SpeechPrune
    & 0.00
    & 58.45
    & 21.04
    & 26.50
    & 62.21
    & 99.04
    & 61.31
    & 74.19
    & 39.60$^\dagger$
    & 13.03
    & 96.27
    & 49.63
    & \textbf{50.03}
    & 17.40 \\

    $10\%$ & DART
    & 0.00
    & 58.50
    & 20.89
    & 26.46
    & 62.27
    & 99.15
    & 64.89
    & 75.44
    & \underline{40.65}$^\dagger$
    & 14.60
    & \underline{96.97}
    & \underline{50.74}$^\dagger$
    & 33.63
    & 17.40 \\

    & Frame
    & 0.00
    & 59.22
    & 21.33
    & 26.85
    & 64.18
    & \textbf{100.00}
    & 61.45
    & 75.21
    & 35.83$^\dagger$
    & 11.79
    & 95.17
    & 47.60
    & 39.17
    & 17.35 \\

    & HeadRouter
    & \underline{5.15}
    & 64.81
    & 19.24
    & 29.73
    & 66.97
    & \textbf{100.00}
    & 66.54
    & 77.84
    & 31.93
    & 13.08
    & 95.39
    & 46.80
    & \underline{40.16}
    & 17.35 \\

    & AudioScorer
    & \textbf{31.25}
    & \textbf{74.57}
    & \textbf{25.14}$^\dagger$
    & \textbf{43.65}
    & \underline{67.82}
    & \textbf{100.00}
    & \textbf{71.24}$^\dagger$
    & \textbf{79.69}
    & \textbf{43.75}$^\dagger$
    & \underline{14.67}
    & \textbf{97.19}
    & \textbf{51.87}$^\dagger$
    & 32.97
    & 17.51 \\
    \hline
  \end{tabular}}

  \caption{AudioMarathon results on Qwen2.5-Omni-3B:
  WAR for ASR and Macro-F1 otherwise. Avg.\ is the unweighted mean
  of each three-task category; Audio Authenticity contains only SD.
  Single/double vertical rules separate category averages/categories.
  \textbf{Bold}/\underline{underlining} mark the best/second-best pruning
  scores per retention ratio; $^\dagger$ marks scores above unpruned.
  Prefill compute: mean TFLOPs to the first token over the same $80$ randomly sampled duration-stratified clips.
  ASR: automatic speech recognition; SCR: speech content reasoning;
  SER: speech entity recognition; SED: sound event detection;
  MC: music classification; ASC: audio scene classification;
  ER: emotion recognition; SAR: speaker age recognition;
  SGR: speaker gender recognition; SD: deepfake detection.}
  \label{tab:am}
\end{table*}

\paragraph{LALM-guided relevance distillation (H2).}
For each training example, we feed the full audio and question to the frozen
LALM, teacher-force the reference answer, and backpropagate its log-likelihood
to the audio tokens at the decoder input. Each token's gradient norm measures
the local sensitivity of the answer to that token. We normalize these values
into soft token-level targets that supervise AudioScorer's ranking. We
complement this with downstream answer supervision. We scale each audio token
by its predicted score: higher-scoring tokens are suppressed less than
lower-scoring tokens. At inference, soft weighting is replaced
by hard Top-$K$ pruning after layer $\ell_p$. Full details are provided in
Appendix~\ref{app:method_details}.

\section{Experiments}
\label{sec:experiments}

\paragraph{Models and evaluation.}
We evaluate AudioScorer on Qwen2.5-Omni-3B~\cite{Qwen2.5-Omni} using AudioMarathon~\cite{he2025audiomarathon} and
MMAU-Pro~\cite{kumar2026mmau}. We compare against the following baselines: FastV~\cite{chen2024image},
DART~\cite{wen2025stop}, uniform frame sampling~\cite{he2025audiomarathon},
HeadRouter~\cite{he2026headrouter}, and SpeechPrune~\cite{lin2025speechprune}
at matched $70\%$, $40\%$, and $10\%$ retention. All methods prune after
decoder layer $\ell_p=2$, so the first two layers process the full sequence.
The same ranking is reused across retention ratios.
\label{sec:results_qwen}

% =====================================================================
% AudioMarathon
% =====================================================================

% =====================================================================
% MMAU-Pro overall
% =====================================================================

\begin{table}[t]
  \centering
  \footnotesize
  \setlength{\tabcolsep}{5pt}
  \begin{tabular}{lccc}
    \hline
    \textbf{Method} & $70\%$ & $40\%$ & $10\%$ \\
    \hline
    Unpruned & 48.57 & 48.57 & 48.57 \\
    \hline
    FastV & 48.53 & 48.78$^\dagger$ & \underline{47.17} \\
    SpeechPrune & 48.14 & 47.86 & 46.01 \\
    DART & 48.45 & 47.90 & 46.27 \\
    Frame & 48.42 & 48.20 & 46.41 \\
    HeadRouter & \underline{48.64}$^\dagger$ & \underline{49.00}$^\dagger$ & 46.16 \\
    AudioScorer & \textbf{49.18}$^\dagger$ & \textbf{49.53}$^\dagger$ & \textbf{48.51} \\
    \hline
  \end{tabular}
  \caption{MMAU-Pro official weighted scores at three retention ratios.
  Bold, underlining, and $^\dagger$ follow Table~\ref{tab:am}.}
  \label{tab:mmau_overall}
\end{table}

% =====================================================================
% MMAU-Pro category breakdown
% =====================================================================

% =====================================================================
% Discussion
% =====================================================================

% Queue the full-width ablation table early enough for page four.
% Generated by Results/make_ablation_appendix.py.
\begin{table*}[!t]
  \centering\footnotesize
  \setlength{\tabcolsep}{3pt}
  \begin{tabular}{l|r|r|r|r||r|r|r|r||r|r|r|r}
    \hline
    & \multicolumn{4}{c||}{$70\%$ retention} & \multicolumn{4}{c||}{$40\%$ retention} & \multicolumn{4}{c}{$10\%$ retention} \\
    \cline{2-5}\cline{6-9}\cline{10-13}
    Variant & SC & AC & SI & AA & SC & AC & SI & AA & SC & AC & SI & AA \\
    \hline
    AudioScorer (full) & \textbf{68.49} & 80.16 & \textbf{49.75} & 82.47 & 68.01 & 80.29 & 49.27 & 71.01 & \textbf{43.65} & 79.69 & \textbf{51.87} & 32.97 \\
    w/o question conditioning & 68.13 & 79.95 & 49.55 & 81.06 & \textbf{68.48} & 80.31 & 49.67 & 64.76 & 41.84 & 80.26 & 51.22 & 31.36 \\
    w/o relevance distillation & 67.90 & 79.93 & 49.70 & 38.92 & 63.04 & 80.30 & \textbf{50.04} & 35.64 & 27.72 & 73.12 & 48.41 & 44.03 \\
    w/o both & 66.92 & \textbf{80.83} & 49.74 & 81.23 & 64.96 & 80.41 & 41.44 & 43.98 & 27.58 & 75.42 & 34.15 & \textbf{54.54} \\
    w/o downstream CE & 68.33 & 80.38 & 49.53 & 84.13 & 68.29 & \textbf{80.69} & 49.01 & \textbf{79.64} & 31.20 & 74.02 & 51.71 & 51.06 \\
    Single-layer scoring head & 68.16 & 80.50 & 49.17 & 82.21 & \textbf{68.48} & 80.42 & 49.12 & 67.81 & 35.74 & \textbf{80.50} & 51.03 & 32.35 \\
    Hard Top-$K$ + STE & \textbf{68.49} & 80.07 & 49.22 & \textbf{84.52} & 68.30 & 80.40 & 49.77 & 71.29 & 34.46 & 75.15 & 51.61 & 52.71 \\
    \hline
  \end{tabular}
\caption{AudioMarathon ablations by category: SC = Speech Content,
AC = Audio Classification, SI = Speaker Information, and
AA = Audio Authenticity. Scores are unweighted task means;
\textbf{bold} marks the best result in each column.
The joint ablation (w/o both) removes question conditioning via
cross-attention and relevance distillation.
CE denotes cross-entropy. STE denotes the straight-through estimator~\cite{bengio2013estimating}.}
  \label{tab:ablation}
\end{table*}

\subsection{Results}
\label{sec:results}
\begin{figure}[t]
  \centering
  \includegraphics[width=\linewidth]{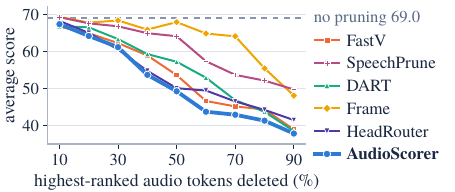}
  \caption{Deleting the top-ranked audio tokens on $500$ random equally drawn AudioMarathon samples: a steeper drop indicates better evidence localization.}
  \label{fig:deletion}
\end{figure}

\begin{figure}[t]
  \centering
  \includegraphics[width=1.0\linewidth]{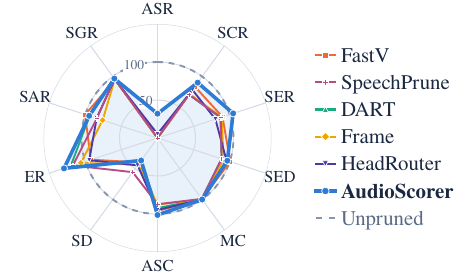}
  \caption{AudioMarathon subtask scores at $10\%$ retention, normalized to the
  unpruned model (WAR for ASR, Macro-F1 otherwise).}
  \label{fig:radar}
\end{figure}

\paragraph{Does learned token scoring preserve answer quality?}
The benefit of AudioScorer is most pronounced on speech tasks under aggressive
compression. At $10\%$ retention, it improves Speech Content by $13.38$ points over the strongest baseline, as shown in Table~\ref{tab:am}. This improvement is substantially larger than those observed for Audio Classification ($0.57$ points) and Speaker Information ($1.13$ points). SpeechPrune leads on Audio Authenticity. AudioScorer is best or tied-best on seven subtasks; its Speech Content advantage includes ASR ($31.25$ versus $5.15$ for HeadRouter) and SCR ($74.57$ versus $69.20$ for FastV). At $40\%$ and $70\%$, it also leads on Speech Content. On MMAU-Pro, it achieves the highest weighted score among pruning methods at all three budgets and, at $10\%$ retention, stays within $0.06$ points of the unpruned model ($48.51$ versus $48.57$), as shown in Table~\ref{tab:mmau_overall}. Appendix~\ref{sec:mmau_details} gives all category results. Figure~\ref{fig:deletion} further shows that deleting each method's highest-ranked tokens degrades AudioScorer most on average, supporting the informativeness of its ranking. AudioScorer's selection is relatively cheap: it adds on average $0.16$ TFLOPs over Frame, which prunes at the same depth without scoring, i.e., $0.4$--$0.9\%$ of total prefilling compute as shown in Table~\ref{tab:am}.

\paragraph{How does pruning sensitivity vary across tasks?}
Figure~\ref{fig:radar} shows strong task-dependent sensitivity to pruning. At $10\%$ retention, ASR nearly collapses for most methods and SD degrades substantially, while MC, SGR, SED, and ASC remain more robust. ER is a notable exception: pruning improves it for most methods, with the largest gain for AudioScorer ($33.82\!\rightarrow\!43.75$). MMAU-Pro shows a similar effect, where $40\%$ retention improves the weighted score from $48.57$ to $49.53$. One possible explanation is that pruning filters irrelevant context and mitigates \emph{lost-in-the-middle} effects~\cite{liu2024lost}, though our results do not establish this as the sole cause.

% \begin{figure}[t]
%   \centering
%   \includegraphics[width=0.62\linewidth]{figures/er_vs_pruning.pdf}
%   \caption{Emotion-recognition performance across pruning ratios.}
%   \label{fig:er_curve}
% \end{figure}

\section{Ablation Study}
\label{sec:ablation}

Table~\ref{tab:ablation} shows that both of our supervision signals are important,
especially under aggressive pruning. Relevance distillation has the largest
effect: removing it substantially degrades Speech Content and Audio
Classification at $10\%$ retention. Removing downstream CE causes a
similar but smaller decline. This supports their complementary roles:
gradient-based targets provide explicit token-level relevance supervision, whereas
CE directly optimizes the retained evidence for answer quality. Simplifying the scoring head to a single
linear layer reduces Speech Content performance at $10\%$ retention. Replacing soft weighting with
hard Top-$K$ training using a straight-through estimator
(STE)~\cite{bengio2013estimating} also hurts Speech Content and Audio
Classification at $10\%$ retention, while remaining competitive or slightly
better at $40\%$ and $70\%$. This suggests that soft weighting is most helpful
when pruning is aggressive. Question cross-attention has a smaller and more task-dependent effect,
suggesting that much of the gain comes from learned relevance supervision, with
explicit question conditioning providing additional adaptation. No component is
uniformly beneficial across all categories. Full subtask results are provided in
Appendix~\ref{sec:ablation_details}.

% Keep all main-paper floats before the back matter and references.
\FloatBarrier

% Intentionally empty (kept so that \input{sections/bibtex_files} resolves).

\section{Related Work}
\label{sec:related}

FastV~\cite{chen2024image} uses attention-based
ranking, DART~\cite{wen2025stop} removes redundant tokens, and
AudioMarathon~\cite{he2025audiomarathon} evaluates these methods for long audio
alongside uniform sampling. SpeechPrune~\cite{lin2025speechprune} combines speech--text similarity with approximated attention, while HeadRouter~\cite{he2026headrouter} uses calibrated head profiles.
Other methods reduce tokens through merging rather than pruning~\cite{xiang2026we, ToMe}. LearnPruner~\cite{takezoe2026learnpruner} learns question-independent vision pruning, introducing question conditioning only later within the decoder. AudioScorer instead learns question-conditioned audio-token relevance from reference-answer gradients and downstream supervision, contributing a new supervision and pruning design for audio.

\section{Conclusion}

AudioScorer learns question-conditioned audio-token relevance using supervision from a frozen LALM. At aggressive retention budgets, it improves speech recognition, audio classification, and speaker information modeling over pruning baselines. The category-level trade-offs motivate further work on task-adaptive selection for efficient long-form audio understanding.

\section*{Limitations}

Our experiments are limited to Qwen2.5-Omni-3B and the audio modality. Moreover, performance gains are not uniform across tasks under aggressive pruning. For example, audio authenticity detection degrades substantially
at $10\%$ retention. Improvements over competing methods should therefore be
interpreted in the context of these task-dependent trade-offs.

Question conditioning yields modest and task-dependent gains in our ablations. Neither our training datasets nor the evaluation benchmarks directly target
question-specific token selection. AudioScorer is trained from individual
audio--question--answer examples, without explicit supervision that contrasts
which evidence should be selected for different questions on the same
recording. Likewise, the benchmarks evaluate downstream task performance
rather than question-dependent changes in the selected evidence. We therefore
cannot determine whether the modest gains reflect a limitation of the
conditioning mechanism or of the used evaluation.
More controlled question-switching data would allow this effect to be assessed
directly. Our evaluation is subject to benchmark limitations. To the
best of our knowledge, AudioMarathon had not been published in a
peer-reviewed venue at the time of submission.

% \input{sections/acknowledgments}


\bibliography{custom}

\appendix

\section{AudioScorer Architecture and Training Details}
\label{app:method_details}

We describe the configuration of the reported Qwen2.5-Omni-3B checkpoint. The LALM, including its audio encoder, remains frozen. We train only the scorer and its question-conditioning components; the gradient teacher described below has no parameters.

\subsection{Training Details}
\label{app:training_data}
The supervised mixture covers speech transcription, dialogue, sound description, music understanding, speaker characteristics, emotion, and audio authenticity. LibriSpeech~\cite{panayotov2015librispeech} supplies transcription and gender-label examples. MultiDialog and IEMOCAP are drawn from the LongAudio-XL resources~\cite{ghosh2026audio,park2024let,busso2008iemocap}; AudioCaps~\cite{kim2019audiocaps} provides sound captions. WavText5K, YouTube-8M, and MusicBench use the AudioSkills-XL resources~\cite{ghosh2026audio,deshmukh2022audio,abu2016youtube,melechovsky2024mustango}. Audio-authenticity examples come from the \texttt{deepfake-audio-detection} dataset~\cite{deepfake_audio_detection_v4_2024}. 
We train for three epochs on a single NVIDIA A6000 GPU using
AdamW~\cite{loshchilov2017decoupled} with a learning rate of $10^{-4}$.
Training AudioScorer takes approximately $9$ GPU-hours.
% FTAR~\cite{ftar} supplies a separate temporal-localization stream.
To balance the training sources, we randomly sample $1{,}000$ examples from each dataset without replacement using a fixed seed of $42$. The selected subsets remain fixed across epochs. Final evaluation uses the available $6{,}347$ AudioMarathon and $5{,}305$ MMAU-Pro examples.
\subsection{Question-Conditioned Scoring}
Let $\mathbf{X}=\{\mathbf{x}_i\}_{i=1}^{N}$ denote the projected audio
tokens. We encode the user question with
\texttt{BAAI/bge-base-en-v1.5}~\cite{bge_embedding}, producing token-level
representations
$\mathbf{T}=\{\mathbf{t}_j\}_{j=1}^{L}$, where $L$ is the number of question
tokens. These representations are projected into the scorer space and used
as keys and values for cross-attention with the audio tokens.

Of the question encoder's $109.48$M parameters, only the final two
transformer layers ($14.18$M) and pooler ($0.59$M) are trainable; the
remaining $94.72$M are frozen. A trainable projection and LayerNorm
($1.58$M parameters) map $\mathbf{T}$ into the scorer space. The complete
question-encoder module contains $111.06$M parameters, of which $16.35$M are
trainable.

A grouped-query cross-attention layer uses audio tokens
as queries and question tokens as keys and values:
\begin{equation}
 \mathbf{C}=\operatorname{CrossAttn}(\mathbf{X},\mathbf{T}),
 \qquad \mathbf{c}_i=\mathbf{C}_i .
 \label{eq:app_xattn}
\end{equation}
The cross-attention layer follows the Qwen2.5-Omni-3B~\cite{Qwen2.5-Omni} configuration,
using hidden size $2048$, $16$ query heads, $2$ key/value heads, and head
dimension $128$.

A scoring head $f$, an MLP with dimensions
$2048\!\rightarrow\!4096\!\rightarrow\!1024\!\rightarrow\!1$, maps the
question-conditioned representation $\mathbf{c}_i$ of each token to a
logit, and a sigmoid turns the logit into the token score:
\begin{equation}
 s_i=\sigma\!\left(f(\mathbf{c}_i)\right)\in(0,1).
 \label{eq:app_score}
\end{equation}
Overall, AudioScorer has $50.15$M trainable parameters, of which $16.35$M
belong to the question encoder and its projection. Together with the
$94.72$M frozen parameters of the question encoder, it contains $144.87$M
parameters. Relative to the approximately $3.72$B parameters of the frozen
audio-language path of Qwen2.5-Omni-3B (language model plus audio
encoder), AudioScorer corresponds to about $3.89\%$ of the base model
size, while its trainable parameters correspond to about $1.35\%$.

\subsection{Training Losses}
AudioScorer is trained with two losses: the answer-based gradient-rank
distillation loss $\mathcal{L}_{\mathrm{rank}}$ and the answer likelihood
loss $\mathcal{L}_{\mathrm{ans}}$. Both are computed from the reference
answer through the frozen LALM, but on different inputs:
$\mathcal{L}_{\mathrm{rank}}$ from a separate pass on the unweighted audio tokens
and $\mathcal{L}_{\mathrm{ans}}$ on the score-weighted audio tokens.

\paragraph{Answer-based gradient-rank distillation loss.}
We first provide AudioScorer with explicit token-level relevance targets
derived from the gradient saliency of the frozen LALM.

Let
$\mathbf{X}=\{\mathbf{x}_i\}_{i=1}^{N}$ denote the sequence of $N$ projected
audio embeddings, with $\mathbf{x}_i\in\mathbb{R}^{D}$ the $i$-th embedding,
$q$ the user question, and
$\mathbf{y}=\{y_m\}_{m=1}^{M}$ the reference answer of $M$ tokens.
For every training example, we run one additional teacher-forced pass of the
frozen LALM on the \emph{unweighted} audio sequence and compute the
log-likelihood of the reference answer,
\begin{equation}
 J=\sum_{m=1}^{M}
 \log p(y_m\mid y_{<m},q,\mathbf{X}),
 \label{eq:app_grad_obj}
\end{equation}
where $y_{<m}$ denotes the reference-answer tokens preceding position $m$.

We measure the saliency of audio token $\mathbf{x}_i$ by the Euclidean norm of the
gradient of $J$ with respect to its embedding at the input of the first
decoder layer:
\begin{equation}
 g_i=
 \left\lVert
 \frac{\partial J}{\partial \mathbf{x}_i}
 \right\rVert_2 ,
 \qquad i=1,\ldots,N.
 \label{eq:app_grad}
\end{equation}
Here, $g_i\in\mathbb{R}_{\geq 0}$ measures the local sensitivity of the
reference-answer likelihood to token $\mathbf{x}_i$; a larger value indicates greater
sensitivity.

Because inference uses hard Top-$K$ pruning, only the relative ordering of
the token scores determines which tokens are retained; their absolute
magnitudes do not affect the selection. We therefore retain only the
within-example ranking induced by $\{g_i\}_{i=1}^{N}$:
\begin{equation}
 u_i=
 \frac{\operatorname{rank}(g_i)}{N-1}
 \in[0,1],
 \label{eq:app_rank}
\end{equation}
where $\operatorname{rank}(g_i)\in\{0,\ldots,N-1\}$ assigns rank $0$ to the
least salient token and rank $N-1$ to the most salient. Thus, $u_i$ is the
normalized saliency rank of token $i$.

We map the normalized rank to a continuous target using a smooth,
monotonically increasing function $\phi$:
\begin{equation}
 \tau_i=\phi(u_i)\in(0,1),
 \label{eq:app_rank_target}
\end{equation}
where $\tau_i$ denotes the distillation target for audio token $i$.
In our experiments, $\phi$ is a sigmoid centered at the median rank,
assigning targets above $0.5$ to the more salient half of the tokens
and below $0.5$ to the less salient half.

The targets supervise AudioScorer's predicted scores $s_i$ through
\begin{equation}
 \mathcal{L}_{\mathrm{rank}}
 =
 \frac{1}{N}\sum_{i=1}^{N}
 \left(s_i-\tau_i\right)^2.
 \label{eq:app_kd}
\end{equation}

The teacher is computed online during training, has no trainable parameters, and requires one
additional forward and backward pass through the frozen LALM per training
example. Because computing $\tau_i$ requires the reference answer, the teacher
is used only during training. At inference, AudioScorer predicts the token
scores directly from the audio and question.

\paragraph{Answer likelihood loss.}
Gradient-rank relevance distillation provides explicit token-level supervision, but
does not directly optimize whether the predicted scores preserve the
information needed to generate the reference answer. We therefore complement
$\mathcal{L}_{\mathrm{rank}}$ with downstream answer supervision.

Using the predicted score $s_i$, we softly weight each projected audio
embedding as
\begin{equation}
 \tilde{\mathbf{x}}_i
 =
 \left[\epsilon+(1-\epsilon)s_i\right]\mathbf{x}_i,
 \qquad
 \epsilon=0.01,
 \label{eq:app_soft}
\end{equation}
where $\epsilon$ is a small residual weight that prevents low-scoring tokens
from being completely removed during training. We denote the resulting
weighted sequence by
$\tilde{\mathbf{X}}=\{\tilde{\mathbf{x}}_i\}_{i=1}^{N}$.

The weighted sequence is passed through the frozen LALM, and we minimize the
negative log-likelihood of the same reference answer:
\begin{equation}
 \mathcal{L}_{\mathrm{ans}}
 =
 -\sum_{m=1}^{M}
 \log p(y_m\mid y_{<m},q,\tilde{\mathbf{X}}).
 \label{eq:app_ce}
\end{equation}
Although the LALM parameters remain frozen, gradients from
$\mathcal{L}_{\mathrm{ans}}$ propagate through it to AudioScorer, encouraging
scores that preserve information useful for answering the question. During
training, soft weighting is applied before the audio enters the LLM, so all
decoder layers process the weighted sequence. At inference, the full sequence
is instead processed through the first $\ell_p$ decoder layers before hard
Top-$K$ pruning is applied.

\paragraph{Overall objective.}
The complete objective sums the answer-based gradient-rank distillation loss
and the answer likelihood loss with equal weight:
\begin{equation}
 \mathcal{L}=
 \mathcal{L}_{\mathrm{rank}}
 +\mathcal{L}_{\mathrm{ans}}.
 \label{eq:app_total}
\end{equation}

% Queue wide result tables early so they can occupy consecutive appendix pages.
\begin{table*}[!t]
  \centering
  \footnotesize
  \setlength{\tabcolsep}{2.5pt}
  \resizebox{\textwidth}{!}{%
\begin{tabular}{l|cccccccccccc||c}    \hline
    \textbf{Method}
    & \textbf{Speech} & \textbf{Sound} & \textbf{Music}
    & \textbf{Voice} & \textbf{Spatial} & \textbf{Multi}
    & \textbf{Open} & \textbf{Instr.}
    & \textbf{So--Mu} & \textbf{Mu--Sp} & \textbf{So--Sp}
    & \textbf{So--Mu--Sp} & \textbf{Avg.} \\
    \hline

    \multicolumn{14}{c}{\textbf{100\% retention}} \\
    \hline
    Unpruned
    & 46.80 & 41.07 & 55.36 & 42.76 & 46.15 & 38.37
    & 3.210 & 37.93 & 42.00 & 36.96 & 32.95 & 57.14
    & 48.57 \\
    \hline\hline

    \multicolumn{14}{c}{\textbf{70\% retention}} \\
    \hline
    FastV
    & 46.58 & 40.88 & 55.08 & \textbf{44.83}$^\dagger$
    & \textbf{45.85} & \underline{37.91} & 3.210
    & 39.08$^\dagger$ & \textbf{46.00}$^\dagger$
    & \underline{36.96} & 32.95 & \underline{57.14}
    & 48.53 \\

    SpeechPrune
    & 45.45 & \textbf{42.41}$^\dagger$ & 55.57$^\dagger$
    & 39.31 & 43.69 & 35.35 & \underline{3.220}$^\dagger$
    & 37.93 & 40.00 & \underline{36.96}
    & \textbf{36.36}$^\dagger$ & \underline{57.14}
    & 48.14 \\

    DART
    & 45.57 & 41.36$^\dagger$ & 55.57$^\dagger$
    & 43.10$^\dagger$ & 44.62 & \underline{37.91}
    & \textbf{3.247}$^\dagger$ & \underline{40.23}$^\dagger$
    & 42.00 & \underline{36.96} & 29.55
    & \textbf{71.43}$^\dagger$ & 48.45 \\

    Frame
    & \underline{47.03}$^\dagger$ & 41.36$^\dagger$
    & \textbf{56.21}$^\dagger$ & 41.03 & 40.92 & 37.21
    & 3.184 & \underline{40.23}$^\dagger$
    & \underline{44.00}$^\dagger$
    & \textbf{39.13}$^\dagger$ & 32.95
    & \textbf{71.43}$^\dagger$ & 48.42 \\

    HeadRouter
    & 46.80 & 40.78 & \underline{55.99}$^\dagger$
    & \underline{43.45}$^\dagger$ & \underline{44.92}
    & 37.44 & 3.189 & \textbf{43.68}$^\dagger$
    & 40.00 & \underline{36.96}
    & \underline{35.23}$^\dagger$ & \underline{57.14}
    & \underline{48.64}$^\dagger$ \\

    AudioScorer
    & \textbf{48.71}$^\dagger$ & \underline{42.22}$^\dagger$
    & 55.71$^\dagger$ & 43.10$^\dagger$ & 42.77
    & \textbf{40.23}$^\dagger$ & 3.194
    & \underline{40.23}$^\dagger$
    & \underline{44.00}$^\dagger$ & 32.61
    & 34.09$^\dagger$ & \underline{57.14}
    & \textbf{49.18}$^\dagger$ \\
    \hline\hline

    \multicolumn{14}{c}{\textbf{40\% retention}} \\
    \hline
    FastV
    & 46.46 & 42.79$^\dagger$ & \underline{55.78}$^\dagger$
    & 43.10$^\dagger$ & \textbf{44.92} & 34.88
    & 3.204 & \underline{43.68}$^\dagger$
    & \underline{44.00}$^\dagger$ & \underline{36.96}
    & \textbf{36.36}$^\dagger$ & 57.14
    & 48.78$^\dagger$ \\

    SpeechPrune
    & 43.10 & \textbf{44.41}$^\dagger$ & 55.29
    & 38.62 & 42.77 & 34.88 & 3.194 & 37.93
    & 42.00 & \underline{36.96}
    & \underline{34.09}$^\dagger$ & 57.14
    & 47.86 \\

    DART
    & 45.01 & 43.55$^\dagger$ & 55.71$^\dagger$
    & 43.10$^\dagger$ & 37.54 & 35.12
    & \textbf{3.221}$^\dagger$ & 28.74
    & 36.00 & 34.78 & 32.95
    & \underline{71.43}$^\dagger$ & 47.90 \\

    Frame
    & 45.45 & \underline{44.03}$^\dagger$
    & 55.50$^\dagger$ & 42.07 & 40.62 & 34.88
    & 3.164 & 36.78 & 40.00 & \underline{36.96}
    & 32.95 & \textbf{85.71}$^\dagger$
    & 48.20 \\

    HeadRouter
    & \underline{46.58} & 42.98$^\dagger$
    & \textbf{56.28}$^\dagger$
    & \textbf{46.55}$^\dagger$ & \underline{43.38}
    & \underline{35.58} & \underline{3.209} & 37.93
    & \underline{44.00}$^\dagger$
    & \textbf{39.13}$^\dagger$ & 32.95 & 57.14
    & \underline{49.00}$^\dagger$ \\

    AudioScorer
    & \textbf{49.72}$^\dagger$ & 43.17$^\dagger$
    & 55.36 & \underline{45.17}$^\dagger$
    & 40.31 & \textbf{39.53}$^\dagger$ & 3.186
    & \textbf{47.13}$^\dagger$
    & \textbf{46.00}$^\dagger$ & \underline{36.96}
    & \textbf{36.36}$^\dagger$ & 57.14
    & \textbf{49.53}$^\dagger$ \\
    \hline\hline

    \multicolumn{14}{c}{\textbf{10\% retention}} \\
    \hline
    FastV
    & \underline{43.88} & 43.84$^\dagger$
    & \textbf{54.72} & 37.93 & \underline{43.08}
    & 31.63 & 3.191 & \underline{26.44}
    & \textbf{38.00} & \textbf{41.30}$^\dagger$
    & 28.41 & \underline{71.43}$^\dagger$
    & \underline{47.17} \\

    SpeechPrune
    & 38.16 & 45.94$^\dagger$ & 53.10 & 37.24
    & 42.77 & \underline{33.26} & 3.164 & 21.84
    & 32.00 & \underline{36.96} & 28.41 & 57.14
    & 46.01 \\

    DART
    & 38.27 & 46.13$^\dagger$ & 52.40
    & \textbf{40.69} & \underline{43.08}
    & \underline{33.26} & \textbf{3.248}$^\dagger$
    & 21.84 & \underline{34.00} & 28.26
    & 30.68 & 57.14 & 46.27 \\

    Frame
    & 39.51 & \underline{46.61}$^\dagger$
    & 52.82 & 38.62 & \textbf{44.00}
    & 32.56 & 3.173 & 19.54 & 30.00 & 30.43
    & \underline{32.95} & \textbf{85.71}$^\dagger$
    & 46.41 \\

    HeadRouter
    & 40.18 & 45.56$^\dagger$ & 53.31
    & 39.66 & 42.77 & 32.09 & 3.148
    & 17.24 & 28.00 & 26.09 & 28.41
    & \textbf{85.71}$^\dagger$ & 46.16 \\

    AudioScorer
    & \textbf{46.35} & \textbf{46.80}$^\dagger$
    & \underline{54.37} & \underline{40.34}
    & 40.92 & \textbf{33.95} & \underline{3.208}
    & \textbf{34.48} & \textbf{38.00}
    & 34.78 & \textbf{37.50}$^\dagger$
    & 57.14 & \textbf{48.51} \\
    \hline
  \end{tabular}}

  \caption{Detailed MMAU-Pro results at all retention ratios.
  Avg.\ is the official MMAU-Pro weighted~\cite{kumar2026mmau} score across all categories,
  as in Table~\ref{tab:mmau_overall}. Open-Ended QA is scored from
  $0$ to $5$. Mixed-domain abbreviations are So = Sound,
  Mu = Music, and Sp = Speech.
  Bold and underlining denote the best and second-best distinct
  pruning results within each retention ratio;
  $^\dagger$ denotes scores above the corresponding unpruned result.}
  \label{tab:mmau_categories}
\end{table*}

% Generated by Results/make_ablation_appendix.py; category means from logged tasks.
\begin{table*}[!t]
  \centering
  \footnotesize
  \setlength{\tabcolsep}{1.6pt}
  \resizebox{\textwidth}{!}{%
  \begin{tabular}{l|rrrr||rrrr||rrrr||r}
    \hline
    & \multicolumn{4}{c||}{Speech Content} & \multicolumn{4}{c||}{Audio Classification} & \multicolumn{4}{c||}{Speaker Information} & Audio \\
    & \multicolumn{4}{c||}{} & \multicolumn{4}{c||}{} & \multicolumn{4}{c||}{} & auth. \\
    \cline{2-5}\cline{6-9}\cline{10-13}\cline{14-14}
    Variant & ASR & SCR & SER & Avg. & SED & MC & ASC & Avg. & ER & SAR & SGR & Avg. & SD \\
    \hline
    \multicolumn{14}{c}{\textbf{70\% retention}} \\
    \hline
    AudioScorer (full) & 97.49 & 82.99 & \textbf{25.00} & \textbf{68.49} & 69.45 & \textbf{100.00} & 71.03 & 80.16 & \underline{34.67} & 15.09 & \textbf{99.49} & \textbf{49.75} & 82.47 \\
    w/o question-conditioning & 97.42 & 82.88 & 24.08 & 68.13 & 68.24 & \textbf{100.00} & 71.60 & 79.95 & 34.61 & 14.62 & \underline{99.41} & 49.55 & 81.06 \\
    w/o relevance distillation & 97.41 & 83.55 & 22.75 & 67.90 & 68.66 & \textbf{100.00} & 71.13 & 79.93 & \textbf{35.04} & 15.15 & 98.91 & 49.70 & 38.92 \\
    w/o both & \textbf{97.66} & \textbf{83.68} & 19.43 & 66.92 & \textbf{70.60} & \textbf{100.00} & \textbf{71.88} & \textbf{80.83} & 34.30 & \textbf{16.52} & 98.40 & \underline{49.74} & 81.23 \\
    w/o downstream CE & \underline{97.58} & 82.84 & \underline{24.58} & \underline{68.33} & \underline{69.88} & \textbf{100.00} & 71.26 & 80.38 & 33.63 & 15.56 & \underline{99.41} & 49.53 & \underline{84.13} \\
    Single-layer scoring head & 97.51 & 82.97 & 24.01 & 68.16 & 69.81 & \textbf{100.00} & \underline{71.68} & \underline{80.50} & 32.85 & 15.17 & \textbf{99.49} & 49.17 & 82.21 \\
    Hard Top-$K$ + STE & 97.53 & \underline{83.57} & 24.36 & \textbf{68.49} & 69.35 & \textbf{100.00} & 70.85 & 80.07 & 32.25 & \underline{15.92} & \textbf{99.49} & 49.22 & \textbf{84.52} \\
    \hline\hline
    \multicolumn{14}{c}{\textbf{40\% retention}} \\
    \hline
    AudioScorer (full) & \underline{96.94} & \underline{83.25} & 23.85 & 68.01 & 69.10 & \textbf{100.00} & 71.78 & 80.29 & 34.19 & 14.42 & \underline{99.19} & 49.27 & 71.01 \\
    w/o question-conditioning & 96.75 & \textbf{83.54} & 25.16 & \textbf{68.48} & 69.02 & \textbf{100.00} & 71.90 & 80.31 & 34.39 & 15.73 & 98.90 & 49.67 & 64.76 \\
    w/o relevance distillation & 86.97 & 80.16 & 21.99 & 63.04 & \underline{69.54} & \textbf{100.00} & 71.37 & 80.30 & \textbf{37.29} & \underline{15.91} & 96.92 & \textbf{50.04} & 35.64 \\
    w/o both & 96.25 & 80.20 & 18.43 & 64.96 & 68.56 & \textbf{100.00} & \textbf{72.66} & 80.41 & 30.01 & \textbf{16.55} & 77.77 & 41.44 & 43.98 \\
    w/o downstream CE & 96.38 & 83.08 & \textbf{25.42} & 68.29 & \textbf{70.17} & \textbf{100.00} & 71.89 & \textbf{80.69} & 33.81 & 14.03 & \textbf{99.20} & 49.01 & \textbf{79.64} \\
    Single-layer scoring head & 96.90 & 83.18 & 25.36 & \textbf{68.48} & 69.10 & \textbf{100.00} & \underline{72.15} & \underline{80.42} & 33.73 & 14.66 & 98.98 & 49.12 & 67.81 \\
    Hard Top-$K$ + STE & \textbf{96.96} & 82.55 & \underline{25.39} & \underline{68.30} & 69.08 & \textbf{100.00} & 72.13 & 80.40 & \underline{34.86} & 15.48 & 98.98 & \underline{49.77} & \underline{71.29} \\
    \hline\hline
    \multicolumn{14}{c}{\textbf{10\% retention}} \\
    \hline
    AudioScorer (full) & \textbf{31.25} & \underline{74.57} & \textbf{25.14} & \textbf{43.65} & 67.82 & \textbf{100.00} & 71.24 & 79.69 & 43.75 & 14.67 & \underline{97.19} & \textbf{51.87} & 32.97 \\
    w/o question-conditioning & \underline{28.33} & \textbf{75.17} & \underline{22.03} & \underline{41.84} & \underline{68.67} & \textbf{100.00} & \underline{72.11} & \underline{80.26} & 42.15 & 14.95 & 96.56 & 51.22 & 31.36 \\
    w/o relevance distillation & 0.00 & 61.89 & 21.27 & 27.72 & 62.93 & 90.69 & 65.74 & 73.12 & \textbf{45.44} & \textbf{15.70} & 84.08 & 48.41 & 44.03 \\
    w/o both & 0.00 & 61.23 & 21.51 & 27.58 & 68.60 & 89.52 & 68.15 & 75.42 & 23.82 & 12.10 & 66.53 & 34.15 & \textbf{54.54} \\
    w/o downstream CE & 1.65 & 70.72 & 21.24 & 31.20 & 66.54 & \underline{99.26} & 56.27 & 74.02 & 43.28 & 12.86 & \textbf{98.98} & \underline{51.71} & 51.06 \\
    Single-layer scoring head & 13.19 & 73.84 & 20.20 & 35.74 & 68.60 & \textbf{100.00} & \textbf{72.89} & \textbf{80.50} & 43.26 & 13.75 & 96.07 & 51.03 & 32.35 \\
    Hard Top-$K$ + STE & 16.89 & 65.46 & 21.02 & 34.46 & \textbf{69.12} & 96.02 & 60.31 & 75.15 & \underline{44.00} & \underline{15.31} & 95.52 & 51.61 & \underline{52.71} \\
    \hline
  \end{tabular}}
  \caption{Detailed AudioMarathon ablations on Qwen2.5-Omni-3B.
  ASR uses WAR; other subtasks use Macro-F1.
  Avg. is the unweighted mean within each three-task category;
  Audio Authenticity contains only SD. Task abbreviations follow Table~\ref{tab:am}.
  \textbf{Bold}/\underline{underlining} mark the best/second-best scores per
  retention ratio; ties receive the same formatting. The joint ablation (w/o both)
  removes cross-attention with the question and gradient relevance distillation.}
  \label{tab:ablation_complete}
\end{table*}

\section{SpeechPrune Baseline Implementation}
\label{app:baseline_details}

% Some baselines require model- or budget-specific adaptations to evaluate
% them under the same token-retention ratios as AudioScorer. We describe
% these implementation choices below.

% \subsection{SpeechPrune Reimplementation}
\label{app:speechprune}

We reimplement SpeechPrune~\cite{lin2025speechprune} while preserving its
training-free, two-stage selection procedure. The first stage ranks
temporal audio groups using speech--question cosine similarity, while the
second stage further filters the retained representation using
first-layer attention.

The original method uses a fixed intermediate budget of $750$ tokens.
Because our evaluation compares all pruning methods at matched final token
budgets, we replace this fixed value with a sample-dependent intermediate
budget. Given the requested final number of retained tokens, we divide the
required token removals approximately equally between the similarity and
attention stages. 

For the attention stage, we interpret
token importance as the mean received attention across query positions and
attention heads. After both stages, the selected tokens are restored to their original temporal order and removed tokens are discarded after decoder layer $\ell_p=2$, the same pruning depth as all other methods.

\section{MMAU-Pro Results}
\label{sec:mmau_details}

Table~\ref{tab:mmau_categories} reports detailed MMAU-Pro results
across all categories and retention ratios, including the official weighted
score (Avg.) already reported in Table~\ref{tab:mmau_overall}.

AudioScorer achieves the highest weighted score at all three retention
ratios: $49.18$, $49.53$, and $48.51$ at $70\%$, $40\%$, and $10\%$,
respectively. At $70\%$ and $40\%$ retention, it exceeds both the unpruned
model ($48.57$) and the strongest pruning baseline ($48.64$ and $49.00$).
The largest margin appears at $10\%$ retention, where AudioScorer exceeds
the strongest baseline by $1.34$ points while remaining within $0.06$
points of the unpruned model. It largely preserves Speech performance
($46.35$ vs.\ $46.80$ unpruned), outperforming the next-best method
($43.88$), and leads on Sound, Multi-Audio, and Instruction Following.
This is consistent with AudioMarathon, where its largest advantages
also appear in speech-related tasks under aggressive compression.

The mixed-domain results show additional gains, particularly for
Sound--Music and Sound--Speech at $40\%$ retention and Sound--Speech
at $10\%$. However, these subsets contain only $7$--$88$ examples,
so their differences should be interpreted cautiously. Overall, the gains
remain task-dependent: at $10\%$ retention, Frame performs better on
Spatial, FastV on Music, and DART on Voice.

% \paragraph{Why are MMAU-Pro results more tightly clustered?}
% Differences between pruning methods are smaller on MMAU-Pro than on
% AudioMarathon. MMAU-Pro evaluates a broad range of audio capabilities,
% including acoustic perception, world knowledge, spatial reasoning,
% multi-audio reasoning, and other higher-level skills. Performance can
% therefore depend strongly on the reasoning ability of the frozen LALM
% even when the relevant audio evidence is retained, limiting how much
% token selection alone can affect the final score.

% The aggregate score can further hide category-level differences. Large
% categories such as Music, Sound, and Open-Ended QA are comparatively
% stable across pruning methods, so larger differences on categories such
% as Speech, Voice, Spatial, or Multi-Audio are diluted when aggregated.
% The category-level macro-average in Table~\ref{tab:mmau_categories}
% therefore provides a complementary view to the official weighted score,
% while not replacing it.

% In contrast, AudioMarathon more directly stresses long-form audio
% processing. Several of its tasks, particularly transcription and
% speech-content understanding, depend on retaining relevant evidence from
% long audio sequences. Performance is therefore more sensitive to which
% tokens survive pruning, producing larger differences between selection
% methods.

\section{Additional Ablation Analysis}
\label{sec:ablation_details}

Table~\ref{tab:ablation_complete} reports the complete ablation results
across all ten AudioMarathon subtasks at $70\%$, $40\%$, and $10\%$
retention, revealing effects that are less apparent in the category averages.

\paragraph{Variant definitions.}
The \emph{w/o question conditioning} variant removes scorer cross-attention.
The frozen LALM still receives the question. The \emph{w/o relevance
distillation} variant removes gradient-derived targets but retains
cross-attention and downstream CE. The joint ablation (\emph{w/o both})
removes cross-attention and distillation. \emph{w/o downstream CE} retains
only distillation. The single-layer variant replaces the multilayer scoring head with a linear layer.
STE replaces soft weighting with hard Top-$K$ selection and a
straight-through estimator.

\paragraph{Speech Content subtasks.}
ASR is particularly sensitive to the training objectives under aggressive pruning. At $10\%$ retention, removing relevance distillation reduces ASR from
$31.25$ to $0.00$; removing downstream CE, using a linear scoring head,
and using STE reduce it to $1.65$, $13.19$, and $16.89$, respectively.
SCR is more robust, although removing distillation or using STE still
lowers performance. Removing question cross-attention reduces ASR and
SER by approximately $3$ points each while slightly improving SCR.
At larger budgets, most variants retain near-unpruned ASR performance,
indicating that these differences matter most under severe compression.

\paragraph{Audio Classification and Speaker Information subtasks.}
Within Audio Classification, ASC is particularly sensitive to downstream
CE and the training relaxation. At $10\%$ retention, removing CE reduces
ASC from $71.24$ to $56.27$, while STE lowers it to $60.31$.
Removing relevance distillation instead affects all three subtasks,
including a decline in MC from $100.00$ to $90.69$.
In Speaker Information, the effects are concentrated in specific subtasks.
Removing relevance distillation lowers SGR from $97.19$ to $84.08$ at $10\%$ retention.
Removing both distillation and cross-attention causes larger declines in
SGR ($66.53$) and ER ($23.82$), whereas removing cross-attention alone
has relatively little effect.

% Balance the final page by starting this paragraph in the right column.
\paragraph{Budget-dependent effects.}
At $40\%$ and $70\%$ retention, the linear head and STE often match
or slightly outperform the full model on individual subtasks, including
ASC and SER. Their disadvantages emerge at $10\%$, especially for ASR
and ASC. Removing distillation sharply reduces SD at the larger budgets
but improves it at $10\%$, where STE and removing downstream CE also
help. Component benefits therefore depend on both retention budget and
subtask evidence requirements.

\end{document}